\documentclass[a4paper,11pt]{article}
\usepackage{pos}

\title{The all-particle cosmic ray energy spectrum measured with HAWC}
 \ShortTitle{HAWC CR energy spectrum}

\author*[a]{J. A. Morales-Soto}
\author{J. C. Arteaga-Velázquez$^{a}$}

\forColl{HAWC}

\affiliation[a]{Instituto de Física y Matemáticas,\\ Universidad Michoacana de San Nicolás de Hidalgo, Morelia, Mexico}

\emailAdd{jorge.morales@umich.mx}
\emailAdd{juan.arteaga@umich.mx}

\abstract{Thanks to recent technological development, a new generation of cosmic ray experiments have been developed with more sensitivity to study these particles in the primary energy interval from 10 TeV to 1 PeV, such as HAWC. Due to its design and high altitude, the HAWC gamma-ray and cosmic ray observatory can provide a bridge between the data from direct and indirect cosmic ray detectors. In 2017 the HAWC collaboration published its first result on the total energy spectrum of cosmic rays, which covers the range from 10 to 500 TeV. This work updates the previous result by extending the energy interval of the measured all-particle cosmic-ray energy spectrum up to 1 PeV. The energy spectrum was obtained from the analysis of two years of HAWC's data using an unfolding method. We employed the QGSJET-II-04 model for the energy calibration and the spectrum reconstruction. The results confirm the presence of a knee like feature at tens of TeV, as previously reported by the HAWC collaboration in 2017.}

\FullConference{37$^{\rm{th}}$ International Cosmic Ray Conference (ICRC 2021)\\
		July 12th -- 23rd, 2021\\
		Online -- Berlin, Germany}

\begin{document}
\maketitle

\section{Introduction}

For years the observations of the total spectrum of cosmic rays on the energy interval from 10 TeV to 1 PeV were a challenging task and were mostly dominated by direct cosmic ray experiments such as ATIC-02 \cite{atic_spectra} and CREAM \cite{CREAM_MCsimulations} for E < 100 TeV, and by indirect experiments like ARGO \cite{argo_spectra} and TIBET \cite{tibet} at higher energies. In recent years, new direct and indirect cosmic ray detectors have been developed with advanced instrumentation and improved measuring techniques that will allow to study in detail this energy region. Some of these detectors are NUCLEON \cite{NUCLEON_spectra}, DAMPE \cite{dampe}, GRAPES-3 \cite{grapes}, LHAASO \cite{lhaaso}, and HAWC \cite{HAWC_allparticle}.

In particular, the  High Altitude Water Cherenkov (HAWC) observatory is a dense air shower array with 1,200 photomultipliers (PMTs) installed in 300 water Cherenkov tanks containing a total of 60 ML of water. The Cherenkov detectors are distributed over a flat surface of 22,000 m$^2$. HAWC is located at 4100 m a.s.l. at the Pico de Orizaba Volcano in Puebla, Mexico. One of the main science goals of the HAWC collaboration is to study cosmic rays in the TeV regime.

In 2017, the HAWC collaboration reported its first result on the total spectrum of cosmic rays. It covered the energy region between 10 to 500 TeV and was obtained with 8 months of data \cite{HAWC_allparticle}. The HAWC collaboration reported the existence of a break in the all-particle energy spectrum at (45.7 $\pm$ 1.1) TeV, which has been recently confirmed by NUCLEON \cite{NUCLEON_spectra}. The present study provides results on the all-particle cosmic ray energy spectrum between 10 TeV and 1 PeV with two years of HAWC’s data improving the statistical and systematic uncertainties, and extending the previous energy range up to 1 PeV. Therefore these results provide a link between the measurements from cosmic ray direct and indirect detectors in the TeV energy region. Analyzing the most energetic particles detected by HAWC is crucial to understand the acceleration mechanism, origin and propagation of cosmic rays \cite{CR_sources, CR_sources1, CR_sources3, CR_sources4}. The analysis is based on the Bayes unfolding method \cite{Bayes_unfold, Bayes_unfold2, Bayes_unfold3} applied to HAWC's data. This work is organized in the following way: section \ref{section:data_and_simulation} describes the data and simulations that were employed; the analysis method used for the  reconstruction of the total spectrum of cosmic rays and the results are shown in section \ref{section:analysis}, in section \ref{section:discussion} the results are discussed. Finally, the conclusions are presented in section \ref{section:conclusions}. 

\section{HAWC simulations and data sets}
\label{section:data_and_simulation}

The air shower simulations were made via CORSIKA (v760) \cite{Corsika} with the hadronic interaction models FLUKA (for E < 80 GeV) \cite{Fluka} and QGSJet-II-04 (for E $\geq$ 80 GeV) \cite{Qgsjet}. The interaction between secondary particles and HAWC’s detectors was simulated with GEANT4 \cite{Geant4}.  Eight primary nuclei from protons to iron (H, He, C, O, Ne, Mg, Si and Fe) were simulated following an $E^{-2}$ differential energy spectrum and arrival directions in the range $\theta$ = [$0^{\circ}, 65^{\circ}$] according to the procedure described in \cite{HAWC_allparticle,PoS_HAWC_light_spectrum}. The estimation of the primary energy of the event is performed by a log-likelihood analysis in which the probabilities that the measured lateral distribution of charged particles are produced by protons of different energies are computed and compared \cite{HAWC_allparticle}. The procedure needs a set of four-dimensional tables generated from proton induced air shower simulations. These tables are presented in bins of primary energy, zenith angle, deposited charge $Q_ \mathrm{eff}$, and the radial distance from the shower core to the PMT (see \cite{HAWC_allparticle} for more details). The simulations were weighted according to their mass and energy to model the spectra according to a composition model based on experimental measurements \cite{AMS_MCsimulations, AMS_MCsimulations2, CREAM_MCsimulations, CREAM_MCsimulations2, PAMELA_MCsimulations, HAWC_allparticle}. Some quality cuts were applied to HAWC’s data and Monte Carlo (MC) simulations to diminish the systematic effects in the shower core position and the arrival direction, which are going to be described next. The events must have successfully passed the event reconstruction procedure described in \cite{CRAB_hawc}, must have an arrival direction of $\theta \leq 35^{\circ}$, activated at least 60 channels in a radius of 40 m from the shower core, registered signal in, at least, 75 channels from a total of 1200, and activated more than 30\% of the available channels. Also, the data was restricted to the reconstructed energy interval E = 10$^4$ GeV - 10$^{6}$ GeV. The effects of these cuts were studied by using different values of the aforementioned cuts and by observing the effects on the  systematic errors in the shower core position, energy and arrival direction. It is observed that the systematics are reduced with the selection of the implemented cuts. Also, these cuts do not affect the result of the energy spectrum, this was tested by the reconstruction of the spectrum using simulated data. The angular and shower core position resolution at E = 1 PeV are 0.51$^{\circ}$ and 14.5 m, respectively. The measured data is reconstructed  using the same reconstruction algorithm applied to the simulations. The observation period used for this work was taken from January, 2018 to December, 2019, with a total duration of 703 days. After the event selection, the experimental data set has a total of 1.5$\times$10$^{10}$ showers. 

\section{Reconstruction procedure of the energy spectrum}
\label{section:analysis}

\begin{figure}[!h]
	\begin{center}			
		\begin{minipage}{16cm}
			\centering
			\includegraphics[scale=0.27]{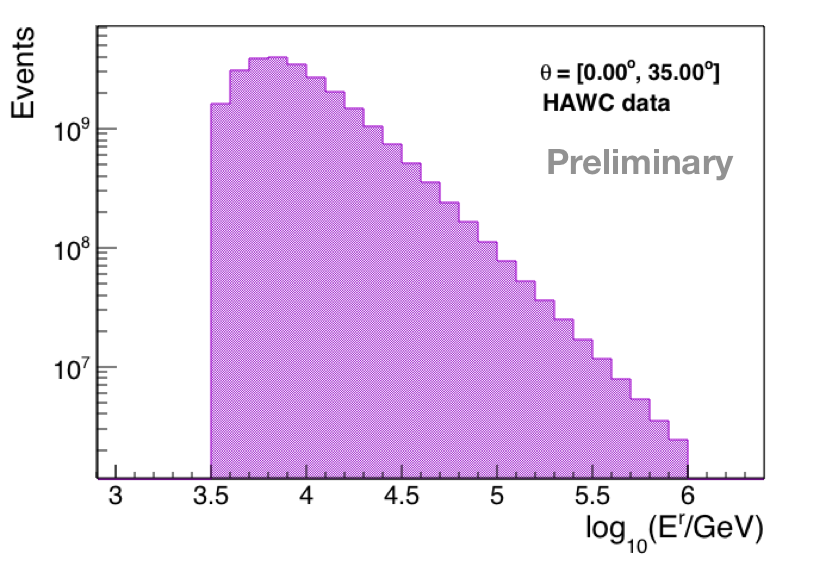}
		\end{minipage}
	\end{center}
	\caption{Raw energy histogram, $N(E^r$), of the selected HAWC data.}
	\label{Fig:raw_histogram}
\end{figure}

\begin{figure}[!h]
	\begin{center}
		\hspace*{-0.4cm}\begin{tabular}{ c c }
		
		\includegraphics[width=0.54\linewidth,height=0.38 \linewidth]{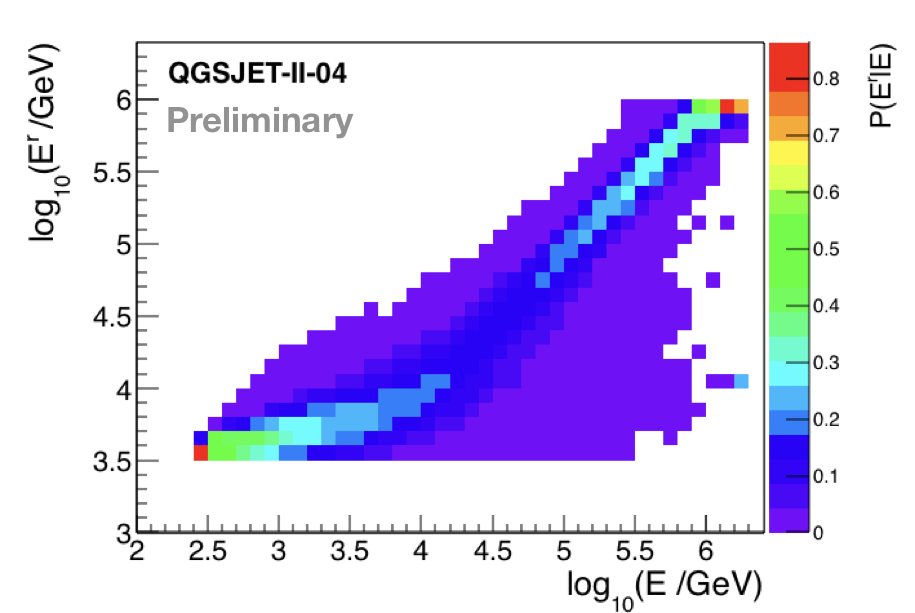}  &
		\includegraphics[width=0.46\linewidth,height=0.38 \linewidth]{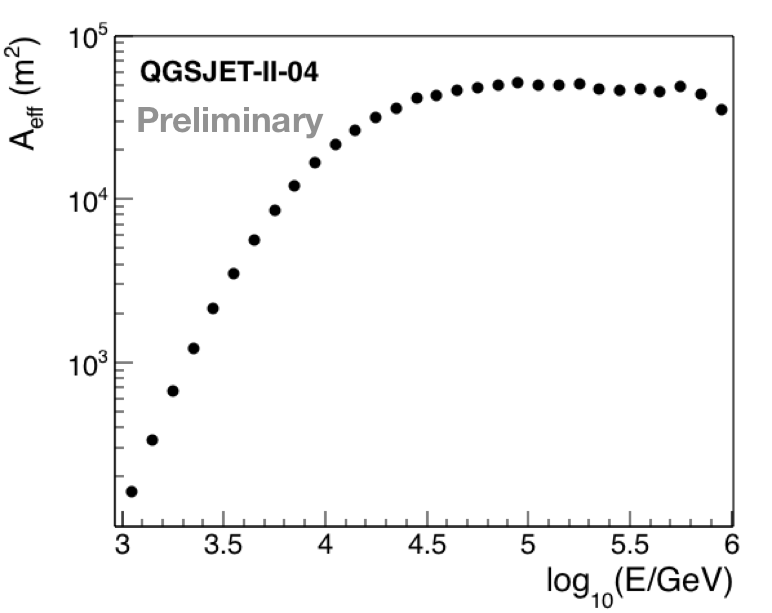}
	\end{tabular}
	\end{center}
	\caption{\textit{Left panel}: The response matrix calculated from simulations. The color palette corresponds to the probability $P(E^{r}|E)$, which takes into account migration effects. The vertical and horizontal axes represent the reconstructed and true shower energies, respectively. \textit{Right panel}: effective area as a function of the primary energy used in the reconstruction of the total energy spectrum. Both, the response matrix and the effective area are obtained from MC simulations with the model already described in section \ref{section:data_and_simulation}.} 
	\label{Fig:response_matrix_eff_area}
\end{figure}

As a first step towards obtaining the all-particle energy spectrum, the energy distribution, $N(E^{r})$, is built from the selected measured data using a bin size of $\Delta \log_{10}$ ($E^{r}$/GeV) = 0.1 (see fig. \ref{Fig:raw_histogram}). However, this energy distribution must be corrected for migration effects in order to find the energy spectrum. For this purpose a Bayes unfolding procedure \cite{Bayes_unfold} is applied employing a response matrix, $P(E^{r}|E)$, which is derived from the MC simulations (see fig. \ref{Fig:response_matrix_eff_area}). 

Once the unfolded energy spectrum, $N(E)$, is obtained, the energy spectrum of cosmic rays is estimated according to the following formula: 

\begin{equation}
    \phi(E) = \frac{N(E)}{ \Delta E \,\, T \,\, \Delta \Omega \,\, A_\mathrm{eff} },
    \label{Eq:Unfolded_spectrum}
\end{equation}

\noindent where $\Delta E$ is the width of the energy bin, $T$ is the effective time of observation, $\Delta \Omega$ is the differential solid angle, and $A_{\mathrm{eff}}$ is the effective area \cite{HAWC_allparticle}:

\begin{equation}
    A_\mathrm{eff}(E) = A_\mathrm{thrown} \cdot \epsilon (E),
    \label{Eq:Effective_area}
\end{equation}

\noindent where $A_{\mathrm{thrown}}$ is the simulated throwing area for the MC events (which is circular) with a radius $R_{\mathrm{thrown} }=1$ km (see \cite{HAWC_allparticle} for further details), and $\epsilon (E)$ is the efficiency for detecting a shower event with energy $E$. The effective area is shown in fig. \ref{Fig:response_matrix_eff_area}, right and it was estimated with MC simulations using our mass composition model.

\subsection{Results}

\begin{figure}[!h]
	\begin{center}
		\hspace*{0cm}\begin{tabular}{ c c }
		
		\includegraphics[width=0.52\linewidth,height=0.38 \linewidth]{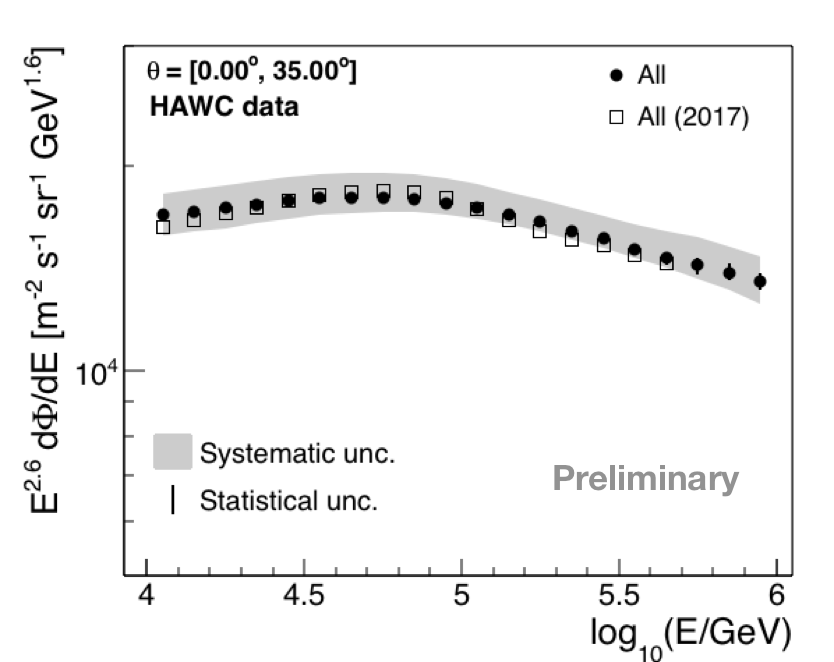}  &
		\includegraphics[width=0.46\linewidth,height=0.38 \linewidth]{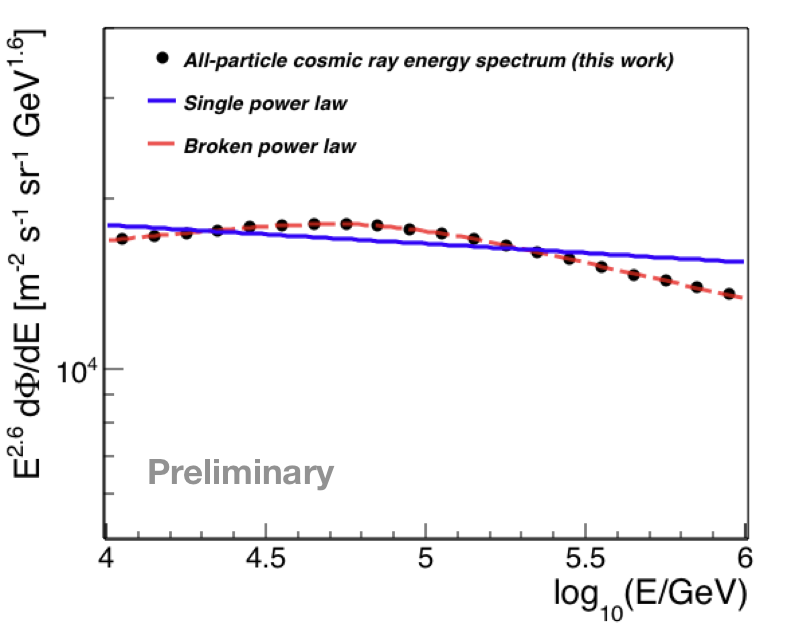}
	\end{tabular}
	\end{center}
	\caption{\textit{Left panel: The unfolded all-particle cosmic-ray energy spectrum obtained from HAWC according to this work (black dots). In here, the energy spectrum is multiplied by an energy factor of $E^{2.6}$. The gray error band corresponds to the systematic uncertainties, while the error bars represent the statistical uncertainties on the flux. For comparison, the measurements of the all-particle cosmic ray energy spectrum presented in \cite{HAWC_allparticle} are shown (open squares). Right panel: Fits to the energy spectrum measured with HAWC (this work) in the energy interval $\log_{10}(E/\mbox{GeV})$ = [4,6]. The blue line represents the fit made with the power law formula from eq. (\ref{Eq:power_law}), while the red dashed line represents the fit made with the broken power law from equation (\ref{Eq:broken_power_law}).}} 
	\label{Fig:HAWC_all_particle_spectrum}
\end{figure}

The unfolded energy spectrum obtained from this analysis is shown in fig. \ref{Fig:HAWC_all_particle_spectrum} left, where the error band represents the systematic uncertainties and the error bars, the statistical errors. In \cite{HAWC_allparticle} the HAWC collaboration reported a break in the spectrum at $E_{knee} = 45.7 \pm 1.1 $ TeV. The spectrum presented in fig. \ref{Fig:HAWC_all_particle_spectrum} left shows also a similar feature at TeV energies. We have fitted the spectrum with a $\chi^2$ procedure using a power-law formula

\begin{equation}
    \Phi(E) = \Phi_{0} E^{\gamma_{1}},
    \label{Eq:power_law}
\end{equation}

\noindent and taking into the account correlation between the data points as described in \cite{PDG}. For the fit, the statistical errors include the uncertainties due to the limited statistics from the data and the response matrix. In equation (\ref{Eq:power_law}), $\Phi_0$ is used as a normalization parameter, and $\gamma_1$ is the spectral index. The results are $\Phi_0 = 24210.3 \pm 332.17$  m$^{-2}$  s$^{-1}$  sr$^{-1}$  GeV$^{-1}$ and $\gamma_1 = -2.63 \, \pm \, 0.01$ with $\chi^{2}_{0}$ = 492.23 for 18 degrees of freedom. Likewise, a $\chi^2$ fit to the spectrum  with a broken power-law function

\begin{equation}
    \Phi (E) = \Phi_{0} E^{\gamma_1} \left[ 1 +   \left( \frac{E}{E_0} \right)^{\epsilon}  \right] ^{(\gamma_2 - \gamma_1)/\epsilon},
    \label{Eq:broken_power_law}
\end{equation}

\noindent yields $\Phi_0 = 10404.95 \pm 97.19$  m$^{-2}$  s$^{-1}$  sr$^{-1}$  GeV$^{-1}$, $\gamma_1 = -2.54 \, \pm \, 0.01$, $\gamma_2 = -2.72 \, \pm \, 0.01$, $\epsilon =  3.04 \, \pm \, 1.51$, and $E_0 = (69.1 \, \pm \, 7.5) $ TeV  with  $\chi^{2}_{1}$ = 0.61 for 15 degrees of freedom. The results of both fits can be seen in fig. \ref{Fig:HAWC_all_particle_spectrum}, right. A test statistic, $TS = \Delta \chi^2 = \chi_{0}^{2} -\chi_{1}^{2}$, is employed to find out which fit best describes the data. For our result we have that $TS_{obs}$ = 491.62. The next step is to generate  toy MC spectra with correlated data points using our covariance matrix with the results of the fit for the power-law model \cite{PDG}. Then, the fits with eqs. (\ref{Eq:power_law}) and (\ref{Eq:broken_power_law}) were repeated and from here, we calculated the distribution of the TS under the hypothesis that the spectrum is best described by a power law. From this distribution it was found that the \emph{p-value} for $TS_{obs}$ is $p \leq 2 \times10^{-6}$, giving the broken power law scenario a significance of 4.6$\sigma$.

\begin{figure}[!h]
	\begin{center}			
		\begin{minipage}{16cm}
			\centering
			\hspace*{-1.3
			cm}\includegraphics[scale=0.28]{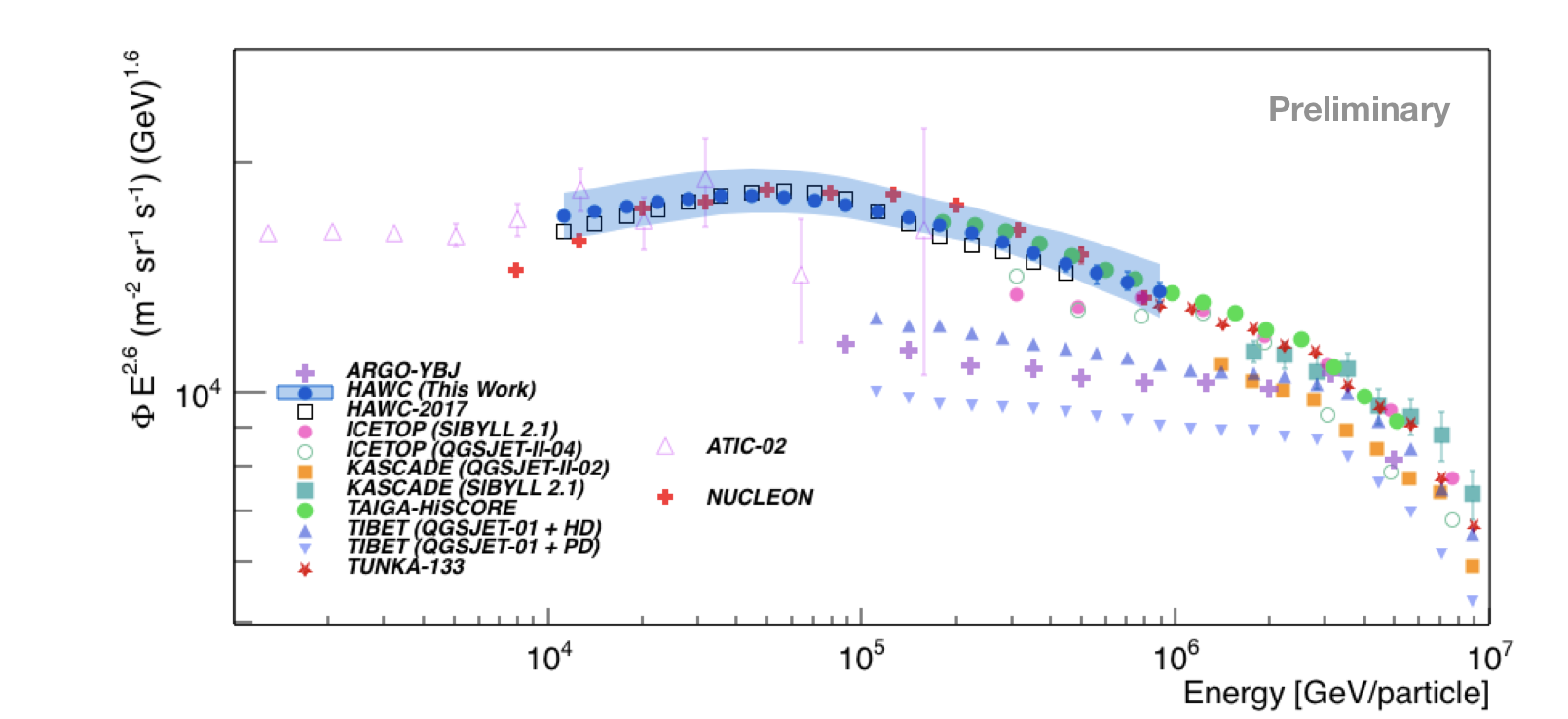}
		\end{minipage}
	\end{center}
	\caption{The all-particle cosmic ray energy spectrum obtained in this work compared with the results from the direct and indirect cosmic ray experiments ATIC-02 (violet open triangles) \cite{atic_spectra}, NUCLEON (red crosses) \cite{NUCLEON_spectra}, ARGO-YBJ (violet crosses) \cite{argo_spectra}, ICETOP (pink dots and green open circles) \cite{ICECUBE_spectra}, KASCADE (orange squares and green squares) \cite{kascade1, kascade2}, TAIGA-HiSCORE (green circles) \cite{taiga}, TIBET (upward blue triangles and downward blue triangles) \cite{tibet}  and TUNKA-133 (red stars) \cite{tunka}. The spectrum is also compared with the previous result from the HAWC collaboration (open blue squares) \cite{HAWC_allparticle}. The error bars and the blue error band correspond to the statistic and systematic uncertainties of the total spectrum from this analysis, respectively, meanwhile the other spectra from the different experiments, and HAWC's previous measurement, are presented only  with their corresponding statistical errors.}
	\label{Fig:comp_spectra}
\end{figure}

The all-particle cosmic ray energy spectrum measured with HAWC (this work) is compared to the results from other direct and indirect experiments in fig. \ref{Fig:comp_spectra}. The measurements are from the satellites ATIC-02 \cite{atic_spectra} and NUCLEON \cite{NUCLEON_spectra}, and from the indirect cosmic ray experiments ARGO-YBJ \cite{argo_spectra}, ICETOP \cite{ICECUBE_spectra}, KASCADE \cite{kascade1, kascade2}, TAIGA-HiSCORE \cite{taiga}, TIBET \cite{tibet}  and TUNKA-133 \cite{tunka}. 

\subsection{Uncertainties in the spectrum}

\begin{figure}[!h]
	\begin{center}			
		\begin{minipage}{16cm}
			\centering
			\hspace*{-1.3
			cm}\includegraphics[scale=0.24]{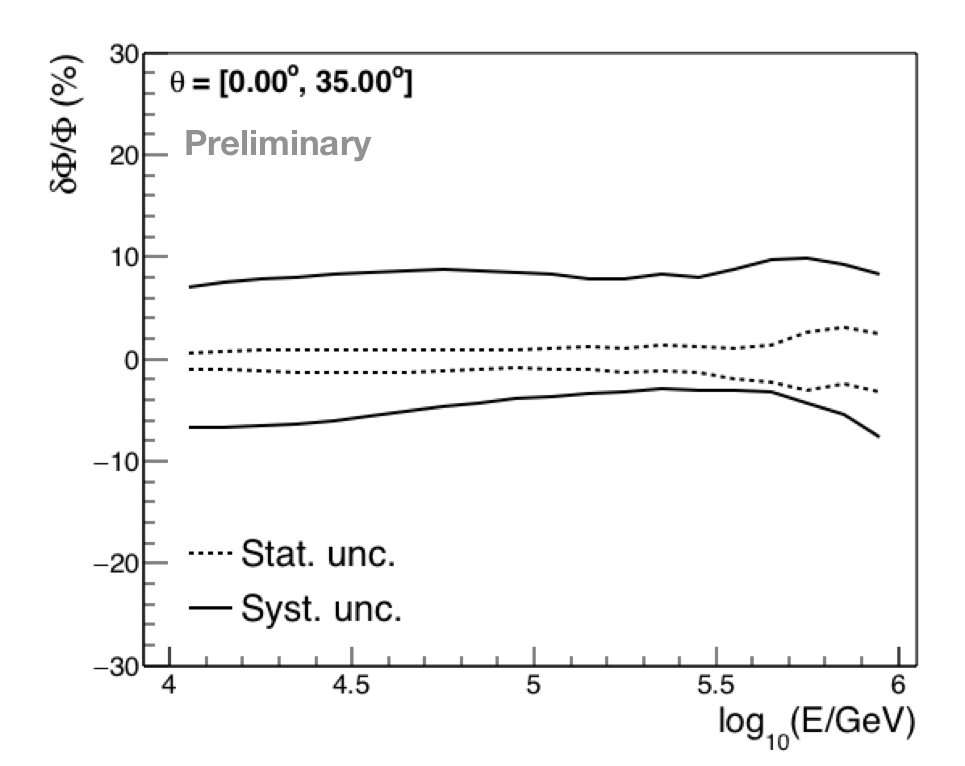}
		\end{minipage}
	\end{center}
	\caption{The systematic (black solid line) and statistical
(slashed line) relative uncertainties vs the primary energy of the all-particle cosmic ray energy spectrum from fig. \ref{Fig:HAWC_all_particle_spectrum}, left.}
	\label{Fig:stat_syst_unc}
\end{figure}

Fig. \ref{Fig:stat_syst_unc} shows the systematic and statistical relative uncertainties, on the energy spectrum obtained in this work as a function of the primary energy. From this figure, it can be observed that at an energy close to E = 1 PeV the statistical uncertainty vary from +2.4\% to -3.2\%. Following the method described in \cite{statistical_errors}, the statistical errors have contributions from the finite size of the experimental data sample and the statistics of the simulations that are used to reconstruct the response matrix. At the same energy the systematic uncertainties are found between +8.3\% and -7.6\%. The sources of systematic errors that were included in this estimation are the bin size, the composition model, the effective area, the quantum efficiency/resolution of the PMTs \cite{CRAB_hawc}, the charge resolution and late light simulation of the PMTs \cite{CRAB_hawc}, the uncertainty of the minimum energy threshold of the PMTs \cite{CRAB_hawc},  and the unfolding technique ( using Gold’s unfolding algorithm \cite{Gold_unfold} in the reconstruction procedure, and including the dependence with the prior regularization procedure).

The systematic errors are dominated by uncertainties on
the PMT performance (+5.1\% to-3.6\%), the 
calculation of the effective area (+4.1\% / -3.5\%), and  the cosmic ray composition model (+5.5\% to -4.9\%).  The later is estimated repeating the reconstruction procedure of the energy spectrum using different composition models: the Polygonato model \cite{Polygonato}, the GSF model \cite{gsf}, and two models derived from fits to measurements from ATIC-2 \cite{atic02} and JACEE \cite{jacee}. The other systematic uncertainties were added in quadrature and together they contribute +5.9\% / -4.3\% to the total systematic uncertainty. 
 
\section{Discussion}
\label{section:discussion}

From the comparison made in fig. \ref{Fig:comp_spectra}, it is observed that there is a good agreement between our result on the all-particle energy spectrum and the measurements from NUCLEON \cite{NUCLEON_spectra} in the interval E = 10 TeV to 1 PeV within systematic errors. At low energies HAWC's data points are also in agreement with the spectrum measured by ATIC-02 \cite{atic02}. On the other hand, HAWC's spectrum is above the measurements from ARGO-YBJ \cite{argo_spectra} and TIBET \cite{tibet}. At high energies HAWC seems to be above the results from ICETOP \cite{ICECUBE_spectra}, close to 1 PeV. Above 100 TeV, HAWC's spectrum is in agreement with the result from TAIGA-HiSCORE \cite{taiga}. The results from this analysis also show a break in the spectrum as reported in \cite{HAWC_allparticle} in the TeV energy region, however, in our result the position of the break is found at higher energies. In \cite{HAWC_allparticle} a sharp break was investigated, and here, a break with some degree of smoothing. In comparison to the result from HAWC in 2017 \cite{HAWC_allparticle}, the systematical uncertainties have been reduced. At an energy of E = 10$^{5}$ GeV, the systematic errors reported in \cite{HAWC_allparticle} are between +26.4\% and -24.8 \%, while the systematic uncertainties from this analysis are found between +8.2\% and -3.6\%. 

\section{Conclusions}
\label{section:conclusions}

We have extended the measurements of the total energy spectrum of cosmic rays with HAWC up to 1 PeV using a data set with high-statistics. In addition to the measurements of NUCLEON \cite{NUCLEON_spectra}, HAWC's result on the all-particle energy spectrum offer a bridge between direct and indirect measurements of the cosmic ray spectrum in the 10 TeV - 1 PeV range. 
The spectrum from this work is in agreement with the measurements from HAWC \cite{HAWC_allparticle}, and the results from NUCLEON \cite{NUCLEON_spectra}. We also confirm the observation of a knee-like structure in the total spectrum of cosmic rays in the TeV energy regime. The position of the break was found at E = (69.1 $\pm$ 7.5) TeV in this study.

\section{Acknowledgments}
\vspace{-0.2cm}
\scriptsize{We acknowledge the support from: the US National Science Foundation (NSF); the US Department of Energy Office of High-Energy Physics; the Laboratory Directed Research and Development (LDRD) program of Los Alamos National Laboratory; Consejo Nacional de Ciencia y Tecnolog\'ia (CONACyT), M\'exico, grants 271051, 232656, 260378, 179588, 254964, 258865, 243290, 132197, A1-S-46288, A1-S-22784, c\'atedras 873, 1563, 341, 323, Red HAWC, M\'exico; DGAPA-UNAM grants IG101320, IN111716-3, IN111419, IA102019, IN110621, IN110521; VIEP-BUAP; PIFI 2012, 2013, PROFOCIE 2014, 2015; the University of Wisconsin Alumni Research Foundation; the Institute of Geophysics, Planetary Physics, and Signatures at Los Alamos National Laboratory; Polish Science Centre grant, DEC-2017/27/B/ST9/02272; Coordinaci\'on de la Investigaci\'on Cient\'ifica de la Universidad Michoacana; Royal Society - Newton Advanced Fellowship 180385; Generalitat Valenciana, grant CIDEGENT/2018/034; Chulalongkorn Universityâ€™s CUniverse (CUAASC) grant; Coordinaci\'on General Acad\'emica e Innovaci\'on (CGAI-UdeG), PRODEP-SEP UDG-CA-499; Institute of Cosmic Ray Research (ICRR), University of Tokyo, H.F. acknowledges support by NASA under award number 80GSFC21M0002. We also acknowledge the significant contributions over many years of Stefan Westerhoff, Gaurang Yodh and Arnulfo Zepeda Dominguez, all deceased members of the HAWC collaboration. Thanks to Scott Delay, Luciano D\'iaz and Eduardo Murrieta for technical support.}
\vspace{-0.2cm}

\setlength{\bibsep}{0pt}

\clearpage
\section*{Full Authors List: \Coll\ Collaboration}

\scriptsize
\noindent
A.U. Abeysekara$^{48}$,
A. Albert$^{21}$,
R. Alfaro$^{14}$,
C. Alvarez$^{41}$,
J.D. \'{A}lvarez$^{40}$,
J.R. Angeles Camacho$^{14}$,
J.C. Arteaga-Vel\'{a}zquez$^{40}$,
K. P. Arunbabu$^{17}$,
D. Avila Rojas$^{14}$,
H.A. Ayala Solares$^{28}$
% ...

\noindent
% ...
$^{14}$Instituto de F\'{i}sica, Universidad Nacional Aut\'{o}noma de M\'{e}xico, Ciudad de M\'{e}xico, M\'{e}xico,
% ...
$^{17}$Instituto de Geof\'{i}sica, Universidad Nacional Aut\'{o}noma de M\'{e}xico, Ciudad de M\'{e}xico, M\'{e}xico,
% ...
$^{21}$Physics Division, Los Alamos National Laboratory, Los Alamos, NM, USA,
% ...
$^{28}$Department of Physics, Pennsylvania State University, University Park, PA, USA,
% ...
$^{40}$Instituto de F\'{i}sica y Matem\'{a}ticas, Universidad Michoacana de San Nicol\'{a}s de Hidalgo, Morelia, Michoac\'{a}n, M\'{e}xico,
$^{41}$FCFM-MCTP, Universidad Aut\'{o}noma de Chiapas, Tuxtla Guti\'{e}rrez, Chiapas, M\'{e}xico,
% ...
$^{48}$Department of Physics and Astronomy, University of Utah, Salt Lake City, UT, USA,
% ...

%
%
%

\end{document}